\documentclass[pdflatex,sn-mathphys-num]{sn-jnl}

\usepackage{graphicx}%
\usepackage{multirow}%
\usepackage{amsmath,amssymb,amsfonts}%
\usepackage{mathrsfs}%
\usepackage[title]{appendix}%
\usepackage{xcolor}%
\usepackage{textcomp}%
\usepackage{manyfoot}%
\usepackage{booktabs}%
\usepackage{algorithm}%
\usepackage{algorithmicx}%
\usepackage{algpseudocode}%
\usepackage{listings}%
 \usepackage{longtable}

 \usepackage{booktabs}
\usepackage{longtable}
 \usepackage{array}
 \usepackage{ragged2e}
 \newcolumntype{L}[1]{>{\RaggedRight\arraybackslash}p{#1}}

 \pdfoutput=1

\makeatletter
\usepackage{enumitem}

\usepackage{amsmath}

\usepackage{tabularx}

\usepackage{booktabs} 

\usepackage{makecell} 
\usepackage{microtype}

\renewcommand\arraystretch{1.15}

\usepackage{tikz}
\usepackage[most]{tcolorbox}

\usepackage{adjustbox}

\usetikzlibrary{positioning,matrix}
\usetikzlibrary{positioning}

\usetikzlibrary{arrows.meta, calc} 

\usepackage{graphicx}

\graphicspath{{figures/}} 

\usepackage{booktabs, tabularx, array, caption, makecell}

\renewcommand{\refname}{References}

\usepackage{array}

\usepackage{booktabs}

\usepackage{ragged2e}
\usepackage{hyperref}
\usepackage{float}
\usetikzlibrary{decorations.pathreplacing}
\renewcommand{\arraystretch}{1.15}

 \newcolumntype{Y}{>{\raggedright\arraybackslash}X}

\usepackage[nameinlink,noabbrev]{cleveref}

\theoremstyle{thmstyleone}
\newtheorem{theorem}{Theorem}

\newtheorem{lemma}{Lemma}
\newtheorem{corollary}{Corollary}

\theoremstyle{thmstyletwo}

\newtheorem{remark}{Remark}

\theoremstyle{thmstylethree}
\newtheorem{definition}{Definition}

\begin{document}

\title[Exact Evolution Law for Action-Weighted Path Ensembles and the Dynamics of Self-Organization]{Exact Evolution Law for Action-Weighted Path Ensembles and the Dynamics of Self-Organization}

\author*[1,2]{\fnm{Georgi} \spfx{Yordanov} \sur{ Georgiev}}\email{ggeorgie@assumption.edu;ggeorgiev@wpi.edu}

\affil[1]{\orgdiv{Department of Biological and Physical Sciences}, \orgname{Assumption University}, \orgaddress{\street{500 Salisbury Street}, \city{Worcester}, \postcode{01609}, \state{MA}, \country{USA}}}

\affil[2]{\orgdiv{Physics Department}, \orgname{Worcester Polytechnic Institute}, \orgaddress{\street{100 Institute Rd}, \city{Worcester}, \postcode{01609}, \state{MA}, \country{USA}}}

\abstract{Self-organizing open systems sustained by source--sink fluxes transform stochastic motion into ordered behavior, yet a general dynamical criterion governing this process has not been established. This paper derives an exact kinematic law for evolving action-weighted canonical path ensembles, and decomposes the dynamics of the ensemble-average action. In the precision-driven regime, the evolution closes: the rate of change is governed by the action variance, so increasing selectivity concentrates probability weight toward lower-action trajectories, causing the average action to decrease monotonically and behave as a Lyapunov-type quantity. Constant and decreasing selectivity lead to stationary and broadening behavior. Endogenous reduction of the average action defines self-organization, while externally prescribed modulation defines controlled ensemble evolution. The framework further admits system-dependent realizations of selectivity dynamics, including reconstruction from observable statistics and feedback-driven evolution, without modifying the underlying kinematic law. By using stochastic action as the organizing trajectory-level quantity, the approach connects path-ensemble organization to stochastic least-action ordering while yielding measurable diagnostics, finite-time constraints, and falsifiable signatures in open stochastic systems.}




\maketitle
\maketitle

\begin{center}
\small
Accepted for publication in \textit{npj Complexity}, on June 23, 2026, in the Collection
``Variational, Nonequilibrium, and Optimization Principles of the
Coevolution of Structure and Dynamics in Complex Systems.''
\end{center}

\section{Introduction}

Understanding why and how open systems increase their degree of organization remains a central problem in nonequilibrium physics. Self-organization in complex systems—from convective flows and chemical oscillations to insect colonies and neural networks—arises from the interplay of stochastic dynamics, dissipation, and feedback across a wide range of physical, chemical and biological systems\citep{prigogine_modern_1967, nicolis_self-organization_1977, cross1993pattern}. Despite its ubiquity, it remains challenging to formulate a quantitative dynamical criterion for organizational growth that can be applied across domains when a canonical path-ensemble representation is available, particularly in transient regimes where feedback reshapes system behavior in time \citep{nigmatullin2021thermodynamic}.

Many proposed signatures of self-organization are expressed in terms of macroscopic observables, including entropy production, free-energy dissipation, mutual information, and order parameters. While these metrics provide deep insight into nonequilibrium steady states, they typically summarize behavior through aggregated state-level statistics rather than describing the continuous evolution of the full ensemble of admissible trajectories. Crucially, their temporal evolution under feedback is not generally constrained to be monotonic; consequently, they do not by themselves establish a formal Lyapunov-type criterion for the emergence of order in open stochastic systems \citep{seifert2012stochastic, sagawa2010generalized, horowitz2014second, ueltzhoffer2021drive, dewar2005maximum}. What has been missing is an exact physical-time law for how a canonical path ensemble itself changes when selectivity and structure evolve. 

From a dynamical perspective, organizational growth can be viewed as the progressive concentration of system behavior onto more efficient, lower-action trajectories within the admissible path space. Particularly in open systems sustained by source--sink fluxes, the trajectory ensemble is the natural object for such a criterion, because organization is expressed not by a single realization but by the redistribution of probability over admissible histories shaped by stochasticity, dissipation, and feedback.

A distinctive feature of the present formulation is that the canonical path ensemble is specifically action-weighted: admissible trajectories are ordered by the stochastic action \(I[\Gamma]\), rather than by a macroscopic order parameter or by another path functional such as entropy production, activity, current, information flow, free energy, or externally prescribed control cost. This choice is not ad hoc. Action is the central trajectory-level variational quantity in mechanics and appears again in path-integral, Onsager--Machlup, and large-deviation descriptions of trajectory probabilities. The present framework uses this action-based ordering to formulate organization as a variational path-space response problem: once admissible histories are assigned stochastic action, the theory describes how the ensemble-average action changes in physical time and identifies conditions under which probability is reweighted toward lower-action trajectories. Thus, the framework extends action-based trajectory weighting from static or asymptotic settings to evolving canonical path ensembles. Other quantities such as entropy, information, dissipation, free energy, or control cost can still enter the formulation when they contribute to the action functional, constrain the path ensemble, or modulate trajectory weights.

In the small-noise limit, path measures concentrate on minimizers, or minimizing sets, of the corresponding action functional, yielding a variational characterization of the limiting dynamics consistent with least-action-type principles \citep{machlup1953fluctuations,freidlin_random_2012,touchette2009large,e_minimum_2004}.  Recent formulations have extended these action-based frameworks to broader classes of nonequilibrium dissipative systems \citep{gay2018variational}, establishing a structural basis for ordering trajectories within canonical ensembles. Within established formulations, however, the ingredients defining the trajectory ensemble—such as the weighting or selectivity parameter, the action structure, and the admissible path space—are specified for a fixed dynamics, an instantaneous inference problem, an asymptotic limit, or an externally prescribed protocol.

The natural next level is the path-ensemble response law for organization. In the endogenous precision-driven regime, this gives a self-organization criterion with a Lyapunov-type branch, a variance-based susceptibility, a reconstruction formula for effective selectivity dynamics, and falsifiability tests. Treating the path ensemble as a dynamical object permits a decomposition of its evolution into distinct structural and selective modes, providing a kinematic way to analyze how changes in selectivity and structure reshape the concentration geometry of the ensemble.

This paper formulates a trajectory-ensemble definition in which self-organization corresponds to endogenous evolution that reduces the ensemble-average action, while externally prescribed modulation corresponds to controlled ensemble evolution. By linking ensemble-level descriptions to measurable dynamics, the formulation provides a testable criterion for self-organization across system-specific realizations and clarifies the distinction between flow-based stochastic processes and static ordered structures. Agent-based models provide a clear setting for illustrating this idea and are widely used to represent biological, social, technological, and networked systems in which local interactions and feedback generate collective organization \citep{camazine2020self,bonabeau2002agent,helbing2001traffic,sumpter2006principles,dorigo2004ant}. In such systems, agent-generated reinforcement of a shared environment can be interpreted as selectivity-modulated organization that amplifies successful trajectories. The contribution of this work is to recast organization as a testable path-ensemble response of an evolving action-weighted trajectory distribution, with reconstruction and falsification available in the precision-driven regime.

\section{Results}

\subsection{Canonical path ensembles as dynamical objects}

While Hamilton's principle identifies extremal trajectories in deterministic settings \citep{Goldstein2002classical, landau_mechanics_1976}, nonequilibrium stochastic systems require a probabilistic treatment of trajectory space, because the relevant object is generally an ensemble of possible histories rather than a single realized path. Path-space theories provide this foundation by assigning statistical weights to complete histories. In particular, Onsager--Machlup, path-integral and large-deviation formulations characterize trajectory probabilities through action-based functionals $I[\Gamma]$, associated with the trajectory $\Gamma$, providing the natural basis for the action-weighted canonical path ensembles considered here \citep{machlup1953fluctuations,feynman2010path,freidlin1998random,freidlin_random_2012,touchette2009large} (see \nameref{MethodsActionFunctional} in Methods). In suitable small-noise or large-selectivity limits, such path measures may concentrate on action-minimizing trajectories, but the general description remains ensemble-level. For a fixed precision/selectivity parameter $\beta$, this static weighting may be written as
\begin{equation}
P[\Gamma] \propto \exp[-\beta I[\Gamma]],
\label{eq:SDLAP_base}
\end{equation}
For \(\beta>0\), lower-action trajectories are assigned larger probability. In the large-precision limit, when such trajectories approach well-defined action minima, the static canonical ensemble concentrates on the corresponding minimizing set, yielding a least-action-type variational characterization \citep{machlup1953fluctuations,freidlin_random_2012,touchette2009large}. This positive-selectivity regime corresponds to the stochastic--dissipative least action principle (SDLAP) \citep{georgiev2025Bio1}; see Methods. Here this structure is extended by writing the action as $I_{\Theta}[\Gamma]$, where $\Theta$ denotes structural parameters of the action landscape (e.g., potential field strengths, dissipation coefficients, or geometric constraints).

Promoting the canonical ensemble specification to a time-dependent object reveals the dynamical evolution of the path ensemble itself. In the core formulation, this occurs through the selectivity parameter \(\beta(t)\) and the structural parameters \(\Theta(t)\); in the most general case, the admissible support \(\Omega(t)\) may also evolve and introduces additional support-dependent terms. In the canonical path ensemble, $\beta(t)$ regulates stochastic dispersion and trajectory preference, appearing formally as an inverse-noise scale or Lagrange multiplier.  The canonical path ensemble is defined by
\begin{equation}
P_t[\Gamma] =
\frac{1}{Z_t}
\exp\!\left[-\beta(t)\, I_{\Theta(t)}[\Gamma]\right],
\qquad
Z_t =
\int_{\Omega(t)}
\exp\!\left[-\beta(t)\, I_{\Theta(t)}[\Gamma]\right]
\mathcal{D}\Gamma,
\label{eq:Canonical_dynamic}
\end{equation}
where \(Z_t\) is the associated partition functional. The reference path measure \(\mathcal{D}\Gamma\) is held fixed, while \(\Omega(t)\) specifies the set of trajectories included in the ensemble at time \(t\). The variable $t$ denotes the physical time indexing the evolution of the trajectory ensemble, which is distinct from the internal time parameter along a trajectory $\Gamma$ that parametrizes the realization itself. Thus, $P_t[\Gamma]$ represents a time-evolving probability measure over complete histories, rather than a distribution over instantaneous states or partial paths. The resulting evolution of \(Z_t\) and the trajectory weights redistributes probability mass across path space, either increasing or decreasing concentration toward lower-action trajectories depending on the balance of selectivity, structural, and support effects.

\subsection{Evolution of the Ensemble-Average Stochastic Action}

Ensemble observables are defined as expectations with respect to the instantaneous path distribution given by the canonical ensemble in~\Cref{eq:Canonical_dynamic}. In particular, the ensemble-average stochastic action is
\begin{equation}
\langle I\rangle_t
=
\int_{\Omega(t)} I_{\Theta(t)}[\Gamma]\,P_t[\Gamma]\,\mathcal{D}\Gamma .
\end{equation}
This quantity serves as the central scalar ensemble-level observable used here to characterize ensemble evolution. It measures the probability-weighted average action of admissible trajectories under the instantaneous path distribution \(P_t\), and therefore tracks how strongly the ensemble is concentrated toward lower-action paths. The following principle characterizes its dynamics. 

\begin{definition}[Average Action Principle (AAP)]
\label{def:SDAAP}
In dynamical canonical path ensembles with time-dependent parameters, the evolution of the ensemble-average stochastic dissipative action is governed by the statistical structure induced by the exponential weighting of trajectories.
\end{definition}

The log-derivative identity for canonical path ensembles (Lemma~\ref{lem:path_weight}, in Methods), together with the regularity conditions specified in \nameref{MethodsAssumptions}, yields the following exact evolution law for the ensemble-average action.

\begin{theorem}[AAP under precision and structural deformation]
\label{Theorem1}
On intervals where the admissible trajectory set $\Omega(t)$ is fixed, the ensemble-average stochastic action satisfies
\begin{equation}
\dot{\langle I\rangle}_t
=
-\dot{\beta}(t)\,\mathrm{Var}_t[I]
+
\left\langle \partial_{\Theta} I_{\Theta} \right\rangle_t \cdot \dot{\Theta}(t)
-
\beta(t)\,\mathrm{Cov}_t\!\big(I_{\Theta}, \partial_{\Theta} I_{\Theta}\big)\cdot \dot{\Theta}(t).
\end{equation}
\end{theorem}

The proof is provided in the Methods (\nameref{Methods:Proof}). This identity constitutes the evolution law for the ensemble-average stochastic action, yielding an exact decomposition of the ensemble dynamics. The selectivity-driven contribution is governed by the variance of the stochastic action, while structural evolution enters through expectation and covariance terms reflecting the dependence of the action on $\Theta(t)$. The covariance term captures the coupling between trajectory reweighting and structural deformation through the precision parameter $\beta(t)$.

\subsection{Scope and validity}

The evolution identity derived in \Cref{Theorem1} applies to systems whose relevant dynamics admit a well-defined canonical path-ensemble representation satisfying the regularity and normalizability conditions stated in \nameref{MethodsAssumptions}  \citep{touchette2009large}. These assumptions ensure that ensemble evolution is well posed and that differentiation under the path integral is valid. The identity is an ordinary differential relation on smooth fixed-support intervals; it does not apply in its present form across discontinuous jumps in \(\beta(t)\) or \(\Theta(t)\), or when the canonical ensemble representation itself breaks down. Support evolution \(\Omega(t)\) requires additional boundary or transport terms. The identity is kinematic rather than mechanistic: it constrains how ensemble observables evolve once \(\beta(t)\) and \(\Theta(t)\) are specified, but it does not determine the system-specific dynamics generating those quantities.

The emphasis in this paper is on canonical path ensembles describing open systems under sustained source--sink flux, as specified in Methods. The formalism itself is broader: in the closed no-flux limit with fixed defining parameters, the canonical path ensemble reduces to stationary equilibrium weighting, whereas closed systems with externally varied parameters remain time-dependent and parametrically driven. The approach is therefore formulated at the path-ensemble level and does not require a particular microscopic stochastic differential equation, a near-equilibrium approximation, or the adoption of entropy production as the fundamental observable.

System-specific dynamics may generate particular evolution laws consistent with this constraint. The kinematic identity yields a local sign classification in the precision-driven regime once the sign of \(\dot{\beta}(t)\) is specified. It does not, by itself, determine the long-time behavior of the canonical path ensemble, such as convergence, saturation, oscillation, or persistent evolution. Such asymptotic behavior requires specification of system-specific closure laws for \(\beta(t)\), and, in the general case, for \(\Theta(t)\) and \(\Omega(t)\). A universal asymptotic classification lies beyond the scope of the present work.

Support evolution, path-space concentration, lower-action concentration, and action-variance reduction are distinct. Support evolution concerns changes in the admissible set \(\Omega\), which specifies which trajectories are included in the ensemble. Path-space concentration concerns how narrowly \(P_t[\Gamma]\) is distributed over admissible trajectories, even when \(\Omega\) itself is fixed. Lower-action concentration concerns whether probability is reweighted toward trajectories with smaller stochastic action \(I[\Gamma]\). Action-variance reduction concerns the dispersion of scalar action values under \(P_t\), as measured by \(\mathrm{Var}_t[I]\). Thus, \(\mathrm{Var}_t[I]\) can change with fixed \(\Omega\), and lower-action concentration need not imply monotonic decrease of \(\mathrm{Var}_t[I]\). The precision-driven identity guarantees decreasing \(\langle I\rangle_t\) for \(\dot{\beta}(t)>0\) and \(\mathrm{Var}_t[I]>0\), but it does not impose a universal monotonic law for the action variance.

\begin{remark}[Evolving admissible sets]
\label{rem:evolvingOmega} 
If the admissible set evolves in time, \(\Omega=\Omega(t)\), the fixed-support derivation is no longer complete, and differentiation of ensemble averages generally produces additional support-dependent contributions. Formally, one may write
\begin{equation}
\dot{\langle I\rangle}_t = -\dot{\beta}(t)\mathrm{Var}_t[I]
+ \langle \partial_{\Theta} I_{\Theta} \rangle_t \cdot \dot{\Theta}(t)
- \beta(t)\mathrm{Cov}_t\!\big(I_{\Theta}, \partial_{\Theta} I_{\Theta}\big)\cdot \dot{\Theta}(t)
+ \mathcal{B}_{\Omega}(t),
\end{equation}
where \(\mathcal{B}_{\Omega}(t)\) represents contributions arising from probability mass entering or leaving the admissible path space as \(\Omega(t)\) changes and additional boundary or transport terms appear. The explicit form of \(\mathcal{B}_{\Omega}(t)\) depends on the rule governing the evolution of admissibility and on how the fixed reference path measure \(\mathcal{D}\Gamma\) weights trajectories near the changing admissible boundary. As a result, support evolution introduces non-universal corrections that cannot, in general, be expressed solely in terms of ensemble covariances. A complete treatment of evolving admissible sets lies beyond the current scope.
\end{remark}

\subsection{Modes of ensemble evolution}

The three deformation modes enter the path-ensemble evolution framework in qualitatively different ways. Selectivity modulation through $\beta(t)$ acts by reweighting trajectories within a fixed action landscape, and in the precision-driven regime its contribution has a sign fixed entirely by $\dot{\beta}(t)$. Structural deformation through \(\Theta(t)\) changes the action functional itself, while support evolution through \(\Omega(t)\) changes the admissible trajectory set, both of which introduce system-dependent contributions that need not have a universal sign. Thus, only the precision-driven component yields a closed variance-controlled regime, whereas structural and support evolution require additional assumptions or system-specific modeling.

\subsection{Precision-driven evolution and variance-controlled organization}

In many physical systems, the parameters governing trajectory weighting and those defining the action functional evolve on different timescales: structural parameters $\Theta(t)$ typically vary slowly, while the selectivity parameter $\beta(t)$ can change rapidly due to feedback or environmental fluctuations \citep{georgiev2025Bio1}. When structural evolution is slow compared to selectivity dynamics, the system operates in an adiabatic (quasi-static) regime in which $\Theta(t)$ may be treated as approximately constant over short intervals. 

Specializing to fast timescales, Theorem~\ref{Theorem1} isolates the precision-driven component of ensemble dynamics. In this regime, the evolution of the ensemble-average action closes exactly.

\begin{corollary}[Precision-driven evolution law]
\label{cor:precision}
Under the conditions of Theorem~\ref{Theorem1}, with fixed structural parameters $\dot{\Theta}(t)=0$, the ensemble-average stochastic action satisfies:
\begin{equation}
\dot{\langle I\rangle}_t = -\dot{\beta}(t)\,\mathrm{Var}_t[I].
\end{equation}
\end{corollary}

This identity establishes that, in the precision-driven regime, the instantaneous rate of change of the ensemble-average action is controlled by the selectivity rate \(\dot{\beta}(t)\) and the ensemble variance \(\mathrm{Var}_t[I]\). The variance \(\mathrm{Var}_t[I]\) acts as a dynamical susceptibility: for a fixed \(\dot{\beta}(t)\), larger variance produces a faster change in the ensemble-average action, while zero variance eliminates this response. In the physical interpretation used here, \(\beta(t)\) is restricted to the nonnegative regime, \(\beta(t)\ge 0\), with \(\beta(t)=0\) corresponding to the absence of action-based selective reweighting.

\subsection{Dynamical regimes and associated structural principles}

The variance-controlled evolution law (\Cref{cor:precision}) induces three dynamical regimes determined by the sign of $\dot{\beta}(t)$. These regimes characterize how changes in selectivity govern the evolution of the ensemble-average action under fixed structural constraints.

\begin{corollary}[Regime classification under precision dynamics]
\label{CorollaryRegimes}
Under the assumptions of Corollary~\ref{cor:precision}, whenever $\mathrm{Var}_t[I]>0$, the relationship between selectivity and average action is defined by the sign identity $\operatorname{sign}(\dot{\langle I \rangle}_t) = -\operatorname{sign}(\dot{\beta}(t))$, resulting in three distinct principles:
\begin{enumerate}[label=\Alph*)]
    \item \textbf{Decreasing Average Action Principle (DAAP):} If $\dot\beta(t)>0$, then $\dot{\langle I\rangle}_t<0$.
    \item \textbf{Stationary Average Action Principle (SAAP):} If $\dot\beta(t)=0$, then $\dot{\langle I\rangle}_t=0$.
    \item \textbf{Increasing Average Action Principle (IAAP):} If $\dot\beta(t)<0$, then $\dot{\langle I\rangle}_t>0$.
\end{enumerate}
\end{corollary}

The regime classification established in \Cref{CorollaryRegimes} provides a sign-level characterization of directional tendencies within the precision-driven regime: depending on the sign of \(\dot{\beta}(t)\), the ensemble-average action decreases, remains stationary, or increases.

The precision-increasing regime, \(\dot{\beta}(t)>0\), corresponds to reweighting probability toward lower-action trajectories. For nonzero action variance, this guarantees monotonic decrease of \(\langle I\rangle_t\), so \(\langle I\rangle_t\) functions as a Lyapunov-type quantity in this branch. The Lyapunov-type property belongs to \(\langle I\rangle_t\), not to \(\mathrm{Var}_t[I]\): the variance controls only the instantaneous susceptibility of \(\langle I\rangle_t\) to selectivity modulation, and its own evolution is not fixed by the precision-driven identity. Thus, \(\mathrm{Var}_t[I]\) need not decrease monotonically; if it increases, the sign of \(\dot{\langle I\rangle}_t\) is unchanged, while the magnitude of the response to a given \(\dot{\beta}(t)\) increases. Decreasing precision, \(\dot{\beta}(t)<0\), produces the opposite reweighting, increasing \(\langle I\rangle_t\) and assigning relatively greater weight to higher-action trajectories.

\begin{definition}[Path-ensemble organization]
\label{def:organization}
Path-ensemble organization refers to the degree to which the realized behavior of a system, represented by its canonical path ensemble, becomes concentrated toward lower-action admissible trajectories. Operationally, organizational increase corresponds to a reduction of the ensemble-average stochastic action \(\langle I\rangle_t\), indicating that system behavior is increasingly weighted toward more action efficient dynamical realizations.
\end{definition}

The condition \(\dot{\langle I\rangle}_t=0\) can arise in two distinct ways within the precision-driven regime: either the selectivity parameter ceases to evolve, so that \(\dot{\beta}(t)=0\), or the action variance vanishes, so that \(\mathrm{Var}_t[I]=0\). These correspond, respectively, to a parameter-stationary regime and an idealized action-degenerate regime in which further selectivity modulation no longer changes the ensemble-average action. The second case is non-generic for physical stochastic systems, where residual fluctuations usually remain. Therefore, within the nondegenerate precision-driven regime, stationarity of \(\langle I\rangle_t\) requires saturation or cessation of the selectivity dynamics, \(\dot{\beta}(t)=0\), while asymptotic approach to stationarity corresponds to \(\dot{\beta}(t)\to0\). Thus, stationarity of the ensemble-average action does not by itself imply full stationarity of the canonical ensemble: in the degenerate limit \(\mathrm{Var}_t[I]=0\), \(\dot{\langle I\rangle}_t=0\) may hold even if \(\dot{\beta}(t)\neq0\). When \(\dot{\beta}(t)=0\), \(\dot{\Theta}(t)=0\), and the admissible support \(\Omega\) is fixed, the canonical measure itself is fixed. In the nondegenerate case \(\mathrm{Var}_t[I]>0\), this recovers the stochastic ensemble setting in which the Least Average Action Principle applies for ensembles with fixed parameters \citep{georgiev2025Bio1}.

\subsection{Self-organization as endogenous reduction of average action}

\begin{definition}[Self-organization]
\label{def:self-org}
Self-organization corresponds to endogenous path-ensemble organization: an internally generated evolution of the canonical path measure that reduces the ensemble-average stochastic action \(\langle I\rangle_t\) through feedback among ensemble statistics and the parameters defining the path ensemble.
\end{definition}

The precision-increasing branch in \Cref{CorollaryRegimes} provides the simplest closed realization of this process: for nonzero action variance, endogenously increasing \(\beta(t)\) monotonically reduces \(\langle I\rangle_t\). Self-organization may also involve endogenous evolution of the structural parameters \(\Theta(t)\) and the admissible set \(\Omega(t)\). When these deformations are coupled to ensemble statistics through internal feedback, they can collectively reduce the ensemble-average action by tuning selectivity, reshaping the action landscape, or constraining the set of admissible trajectories toward more efficient dynamical pathways.

\subsection{Observable consequences of the precision-driven regime}

In the precision-driven regime, observable ensemble statistics provide direct access to the implied precision dynamics and their finite-time and dimensionless observational consequences. Integrating the precision-driven evolution law (\Cref{cor:precision}) over a finite time interval $[t_0, t_1]$ yields the total change in the ensemble-average action:
\begin{equation}
\Delta \langle I\rangle = - \int_{t_0}^{t_1} \dot{\beta}(t)\,\mathrm{Var}_t[I]\,dt.
\label{integral}
\end{equation}
This identity allows both \(\dot{\beta}(t)\) and \(\mathrm{Var}_t[I]\) to vary over the interval; it relates the total change in \(\langle I\rangle_t\) to the time history of selectivity modulation and ensemble fluctuations. Integrating the full fixed-support identity in \Cref{Theorem1} gives an analogous finite-time decomposition with structural contributions, but only the precision-driven term has a sign fixed solely by \(\dot{\beta}(t)\); the structural terms need not have a definite sign.

In experimental settings, \Cref{integral} provides a rigorous interval-level constraint. Given time-resolved measurements of \(\mathrm{Var}_t[I]\) and an independently specified, measured, or proxy-inferred trajectory of \(\beta(t)\), the resulting shift in ensemble-average action is predicted over the interval \([t_0,t_1]\). This enables a direct test of the precision-driven approximation: persistent mismatch between the observed \(\Delta\langle I\rangle\) and the value predicted by \Cref{integral}, beyond statistical uncertainty, indicates the presence of structural deformation \((\dot{\Theta}\neq 0)\), support evolution, or breakdown of the canonical precision-driven description.

To provide a scale-invariant derived observable of organization, the Average Action Efficiency (AAE) \citep{georgiev2025Bio1} is defined in the positive-action regime \(\langle I\rangle_t>0\) as
\begin{equation}
\alpha_t \equiv \frac{\eta}{\langle I\rangle_t},
\label{eq:AAE_def}
\end{equation}
where \(\eta>0\) is a fixed normalization constant with dimensions of action, chosen so that \(\alpha_t\) is dimensionless. In this regime, \(\alpha_t\) is a monotone inverse reparameterization of \(\langle I\rangle_t\). The evolution of $\alpha_t$ follows from \Cref{Theorem1}:
\begin{equation}
\dot{\alpha}_t = \frac{\eta}{\langle I\rangle_t^2} \left[ \dot{\beta}(t)\,\mathrm{Var}_t[I] - \dot{\Theta}(t) \cdot \left( \langle \partial_{\Theta} I_{\Theta} \rangle_t - \beta(t)\,\mathrm{Cov}_t(I_{\Theta}, \partial_{\Theta} I_{\Theta}) \right) \right].
\end{equation}
In the precision-driven regime \((\dot{\Theta}=0)\), the evolution of AAE is governed solely by the selectivity rate and the action variance. For nondegenerate ensembles, \(\mathrm{Var}_t[I]>0\), this reduces to the sign identity:
\begin{equation}
\operatorname{sign}(\dot{\alpha}_t) = \operatorname{sign}(\dot{\beta}(t)).
\end{equation}
Thus, AAE grows when \(\dot{\beta}(t)>0\), is stationary when \(\dot{\beta}(t)=0\), and decays when \(\dot{\beta}(t)<0\).

\subsection{Empirical reconstruction and operational falsifiability}

The ensemble-level evolution law admits empirical reconstruction and yields measurable constraints on observable dynamics. When completed events can be assigned a coarse-grained action cost, \(\langle I\rangle_t\) and \(\mathrm{Var}_t[I]\) can be measured without reconstructing full microscopic trajectories, from aggregate measurements of action per event across an ensemble of realizations \citep{georgiev2025Bio1, georgiev2025Bio2}. In the precision-driven regime, the variance-controlled law (\Cref{cor:precision}) permits reconstruction of the effective selectivity-modulation rate as
\begin{equation}
\dot{\beta}_{\mathrm{rec}}(t) = -\frac{\dot{\langle I\rangle}_t}{\mathrm{Var}_t[I]},
\label{eq:B_growth_identity}
\end{equation}
valid on any interval where \(\mathrm{Var}_t[I] > 0\). Here \(\dot{\beta}_{\mathrm{rec}}(t)\) denotes the rate inferred from measurements of \(\dot{\langle I\rangle}_t\) and \(\mathrm{Var}_t[I]\). If no independent estimate of \(\dot{\beta}(t)\) is available, Eq.~\eqref{eq:B_growth_identity} provides an effective reconstruction of the selectivity dynamics. If \(\dot{\beta}(t)\) is independently specified or estimated---for example through an external protocol, an independently measured proxy, or a system-specific closure---then \(\dot{\beta}_{\mathrm{rec}}(t)\) can be compared with that independent value. Persistent disagreement beyond statistical uncertainty indicates breakdown of the precision-driven approximation and instead points to structural evolution, support changes, or failure of the canonical precision-modulated path-ensemble description. Together, these uses make the precision-driven approximation empirically testable whenever \(\langle I\rangle_t\), \(\mathrm{Var}_t[I]\), and an independent proxy or model for \(\dot{\beta}(t)\) are available.

\subsection{Endogenous precision dynamics and system-specific feedback laws}

In the precision regime  (\Cref{cor:precision}), suppose the $\beta(t)$ evolves according to
\begin{equation}
\dot{\beta}(t)=F\!\left(\langle I\rangle_t,\mathrm{Var}_t[I],\beta(t)\right),
\label{eq:feedback}
\end{equation}
with \(F\) depending on ensemble observables and on the current precision itself. This form allows precision dynamics to be generated endogenously through ensemble-level feedback. The function \(F\) is not specified universally; it represents a system-dependent closure linking ensemble statistics to the evolution of selectivity. 

The canonical measure together with ensemble observables such as $\langle I\rangle_t$, $\mathrm{Var}_t[I]$, and $\beta(t)$ then provides a closed feedback description. In this case, a monotonic reduction of $\langle I\rangle_t$ occurs on intervals where
\begin{equation}
F\!\left(\langle I\rangle_t,\mathrm{Var}_t[I],\beta(t)\right)>0
\quad\text{and}\quad
\mathrm{Var}_t[I]>0.
\end{equation}
Once a system-specific closure \(F\) is specified, the kinematic identity constrains its observable consequences. Combining the feedback law with the variance-controlled evolution law in \Cref{cor:precision} yields a closed-loop system:
\begin{equation}
\dot{\langle I\rangle}_t
=
- F\!\left(\langle I\rangle_t,\mathrm{Var}_t[I],\beta(t)\right)\,\mathrm{Var}_t[I].
\end{equation}
This form makes explicit that, under endogenous feedback, changes in ensemble statistics can in turn modulate trajectory selectivity, thereby closing the loop between dynamics and ensemble structure. Particular functional forms for $F$ can then be chosen to model concrete systems within the general constraints of the framework, as in the example below.

\subsection{Pheromone-mediated foraging as a specific realization}

As a concrete system-specific realization of the generic feedback form in Eq.~\eqref{eq:feedback}, consider pheromone-mediated ant foraging, a standard agent-based model of reinforcement, stigmergic coordination, and path formation in distributed systems in which local agents modify their environment and thereby bias future trajectories. In such systems, source--sink trajectories are reinforced through pheromone deposition, which biases transition probabilities toward previously traversed paths. At the ensemble level, this feedback is represented by an effective precision parameter $\beta(t)$ whose evolution reflects pheromone-mediated reinforcement. In the present simplified coarse-grained description, the admissible trajectory set $\Omega$ and the action functional $I[\Gamma]$ are treated as fixed, so that only the effective selectivity dynamics are modeled explicitly.

The pheromone field is a physical reinforcement field, whereas \(\beta(t)\) is an effective ensemble-level selectivity parameter. The connection between them is indirect: pheromone-mediated reinforcement changes the relative probabilities of trajectories, and this reweighting can be represented at the canonical-ensemble level as modulation of \(\beta(t)\). As reinforcement accumulates along successful paths, probability becomes increasingly concentrated on lower-action, more efficient trajectories. 

A schematic coarse-grained feedback law is
\begin{equation}
\dot{\beta}(t)=\kappa\,\bar{\Phi}_t-\lambda\,\beta(t),
\label{eq:ant_beta}
\end{equation}
where \(\bar{\Phi}_t\) denotes a coarse-grained reinforcement signal associated with successful trajectories and captures the strength of path-dependent bias. The constants \(\kappa>0\) and \(\lambda>0\) set the effective reinforcement gain and the relaxation rate of the selectivity parameter, respectively. The term \(-\lambda\beta(t)\) should be understood phenomenologically: it summarizes the weakening of effective selectivity, for example through pheromone evaporation, environmental noise, or behavioral exploration, without explicitly modeling the microscopic deposition--diffusion--evaporation dynamics of the pheromone field.

Here these processes are compressed into an effective reinforcement signal. When it can be approximated as dependent on the current path ensemble, it may be represented as a functional of the instantaneous canonical measure, \(\bar{\Phi}_t \approx \bar{\Phi}[P_t]\). Thus, the reduced closure is not intended as a microscopic ABM, but as an ensemble-level approximation in which path-dependent reinforcement is summarized by a coarse-grained variable.

In that limit, Eq.~\eqref{eq:ant_beta} defines a system-specific endogenous closure of the general form in Eq.~\eqref{eq:feedback}. The corresponding reduced endogenous loop is
\begin{equation}
P_t \;\mapsto\; \bar{\Phi}[P_t] \;\mapsto\; \dot{\beta}(t) \;\mapsto\; P_t,
\end{equation}

so that, in this reduced description, $P_t$ summarizes the distribution of realized agent trajectories from which the coarse-grained reinforcement signal is computed; this signal modulates selectivity and thereby reshapes the subsequent path ensemble. 

This provides a more detailed system-specific realization of precision dynamics in which the reinforcement field modulates the effective selectivity parameter. Repeated traversal of efficient trajectories increases $\beta(t)$, which further biases the ensemble toward lower-action paths, while the relaxation term $-\lambda\beta(t)$ prevents unbounded growth. Reinforcement and decay can balance, yielding saturation of selectivity at a level set by their interplay. This closes the feedback loop between individual trajectory selection and agent-generated reinforcement of the shared environment, providing a reduced variational description of reinforcement dynamics \citep{camazine2020self, brouillet2024modeling}.

A full agent-based model would include explicitly those effects in the pheromone field.  If the pheromone field develops spatial patterns such as trails or gradients that evolve through deposition, diffusion, and evaporation, then the reduced scalar closure \(\bar{\Phi}_t \approx \bar{\Phi}[P_t]\) is insufficient; the pheromone field should instead be treated as an additional state variable, such as \(\Phi(x,t)\), with its own dynamics. 

While ant foraging provides a canonical example, similar reinforcement-based dynamics arise in traffic flow, collective decision-making, and path-dependent economic dynamics, where effective selectivity emerges from feedback between agent behavior and the shared medium or environment that agents collectively modify \citep{helbing2001traffic,sumpter2006principles,sasaki2018psychology,arthur1987path}. These dynamics also admit a natural representation in network terms, where trajectories correspond to paths on a graph, the action defines path costs, and selectivity modulation biases flow toward lower-cost routes. In this setting, endogenous feedback modifies effective edge weights through accumulated usage, providing a path-space interpretation of reinforcement processes studied in network science \citep{dorigo2004ant,pemantle2007survey}.

\bigskip


\section{Discussion}

The main contribution of this work is the derivation of an exact physical-time evolution law for canonical path ensembles ordered by stochastic action (Theorem~\ref{Theorem1}). This law turns organization into a testable response problem: it constrains how ensemble-level observables respond once a specific parameter evolution is provided. Any environmental or dynamical change that preserves the canonical path-ensemble representation enters through one of the controlling parameters. The identity applies piecewise on intervals over which the admissible support \(\Omega\) is fixed and the parameters \(\beta(t)\) and \(\Theta(t)\) vary smoothly within the same canonical path-ensemble representation. This shift is useful because it makes organizational change calculable as a path-ensemble response: changes in selectivity or structure imply constrained changes in ensemble-average action, variance, and derived efficiency.

The kinematic identity (\Cref{Theorem1}) applies to all canonical path ensembles satisfying the stated regularity and normalizability assumptions on intervals where the admissible support \(\Omega\) is fixed. The interpretation of organization developed here is specifically for open systems with sustained flux between source and sink conditions. In these contexts, the action functional orders complete realizations rather than local equilibrium fluctuations \citep{georgiev2025Bio1}. Consequently, the ensemble describes ongoing processes instead of static configurations, and organizational growth represents the progressive shaping of trajectory statistics under sustained driving. The present results place this notion of organization on an exact ensemble-level footing through the evolution identity in \Cref{Theorem1}. Accordingly, the description of organization shifts from macroscopic configurations or aggregate observables to the underlying evolution of probability distributions over admissible histories that generate them which occurs at the level of probability weights over trajectories, instead of along individual trajectories in state space. In this sense, it may be viewed as a probability flow transverse to trajectories in path space, corresponding to a redistribution of probability mass across trajectories rather than its dynamical motion along them.

Because the identity is kinematic, adaptive environments, endogenous precision dynamics, and structural deformation remain compatible with the formulation whenever their effects can be represented through the parameters defining the canonical path ensemble and the regularity assumptions are maintained. Specific physical mechanisms enter through system-dependent closure laws for these parameters, for example through feedback laws for \(\beta(t)\), rather than by modifying the kinematic identity itself. Evolving admissible support \(\Omega(t)\) can also be considered, but requires additional boundary or transport terms beyond the fixed-support theorem.

The distinction between endogenous and externally prescribed modulation determines whether the ensemble evolution is classified as self-organization or externally controlled. Because \(\langle I\rangle_t\), \(\mathrm{Var}_t[I]\), and the derived AAE can be estimated from coarse-grained action-per-event data, the framework provides a route for reconstruction, numerical testing, and falsification of the precision-driven approximation in systems that admit a canonical path-ensemble representation. Together, empirical reconstruction and system-specific feedback closures link the ensemble-level kinematic law to measurable dynamics.

Compared with approaches tied to particular spatial patterns or system-specific order parameters, a practical strength of the path-space formulation is its domain flexibility. Because the framework is expressed in terms of admissible trajectories and their action-weighted probabilities, it extends in principle to biochemical \citep{nath_coupling_2019}, neural information-processing dynamics \citep{tsai2022path}, and financial time-series systems \citep{marcaccioli2020MaxEnt} whenever these admit a trajectory-level description and a canonical path-ensemble representation \citep{dixit2018MaxCal,bolhuis2021MaxCal}.

Closed or non-source--sink canonical ensembles are not excluded from the formalism, but they lie outside the primary self-organizational interpretation developed here. In such cases, externally prescribed changes in \(\beta(t)\) or \(\Theta(t)\) produce controlled, time-dependent ensemble evolution. Only in the closed no-flux limit with fixed \(\beta\),  \(\Theta\), and $\Omega$, does the canonical ensemble reduce to stationary equilibrium weighting.

The precision-driven regime \Cref{cor:precision} provides the simplest closed realization of path-ensemble organization in the present framework. In the \(\dot{\beta}(t)>0\) branch of \Cref{CorollaryRegimes}, monotonicity is a kinematic consequence of the variance-controlled law, and \(\langle I\rangle_t\) functions as a Lyapunov-type quantity whenever \(\mathrm{Var}_t[I]>0\). In many concrete systems, lower-action concentration may be accompanied by reduced action variance, but this is a system-dependent tendency rather than a universal consequence of the precision-driven identity. Beyond the closed precision-driven branch, structural and support-driven reductions of \(\langle I\rangle_t\) remain possible but system-dependent; whether they admit comparable deformation principles under additional symmetry, conservation, or closure assumptions remains an open problem.

The theory is system-independent at the level of its kinematic path-ensemble identities, but system-dependent at the level of physical realization. In particular, the precision-driven regime yields a closed sign-definite law, whose form does not depend on the microscopic details of the system. What remains system-dependent is the action functional, the admissible path space, and the mechanism or closure that generates \(\beta(t)\).

In the zero-selectivity limit \((\beta=0)\), the canonical weighting becomes uniform over the admissible trajectory set, and the precision-driven mechanism of organization is absent. For the open source--sink systems emphasized here, ensemble change may still arise through structural deformation or support evolution, but not through action-based selective reweighting. A systematic treatment of such structural or constraint-driven organization lies beyond the scope of the present work.

Static or preexisting structures may exhibit order, whether self-organized, externally imposed, or engineered. They enter the present formulation only when they arise from or participate in flow-based stochastic processes that define admissible trajectories and completed action-bearing events. Within this framework, organization at the ensemble level is expressed through the concentration geometry of the path distribution, while physical structure enters through the parameters $\Theta(t)$ and, more generally, through the admissible set $\Omega(t)$ insofar as they shape the action landscape or the space of admissible trajectories and thereby reshape that geometry.

The physical meaning of the precision parameter \(\beta(t)\) is system-dependent: here it should be interpreted as an effective measure of trajectory selectivity. In different realizations, it may correspond to control gains in feedback systems, reinforcement strength in agent-based dynamics, or effective signal-to-noise ratios in driven processes \citep{georgiev2025Bio2}. At fixed structural parameters \(\Theta\), variation in \(\beta(t)\) reflects changes in effective stochastic dispersion or path-selection bias, meaning the relative tendency of the system to favor some admissible trajectories over others. Under endogenous feedback, \(\beta(t)\) evolves because internal variables modulate this selectivity. In empirical settings, it may be inferred indirectly from observable indicators of trajectory consistency or noise suppression, thereby linking the ensemble-level variational principle to measurable dynamics without requiring a universal law for the underlying feedback mechanism \citep{georgiev2025Bio2}.

A stationary canonical path ensemble should not automatically be identified with a physical nonequilibrium steady state (NESS). Here NESS refers, in the open source--sink setting emphasized in this paper, to a physically stationary driven state with sustained fluxes and time-invariant trajectory statistics generated by the underlying dynamics. When \(\beta\), \(\Theta\), and the admissible support \(\Omega\) are fixed, the canonical measure is stationary in the ensemble-level sense: the probability distribution over admissible trajectories no longer changes in physical time. This stationarity corresponds to a physical NESS only if the canonical path measure coincides with, or consistently approximates, the stationary trajectory distribution generated by the underlying dynamics. Establishing such correspondence requires additional assumptions linking the stochastic action functional to the physical process, for example when \(I_\Theta[\Gamma]\) is an Onsager--Machlup functional derived from an underlying stochastic dynamics, or when the canonical ensemble provides a valid coarse-grained approximation to the physical path measure.

The reason for using stochastic action as the organizing observable is that it is not introduced ad hoc. It gives the framework its specific physical and variational basis: action is the central variational quantity of mechanics and appears again in path-integral, Onsager--Machlup, and large-deviation descriptions of trajectory probabilities. In the present framework, this same quantity orders admissible histories in a canonical path ensemble, while the AAP describes how the ensemble-average action changes in physical time. The approach differs from entropy-, information-, free-energy-, guidance-, or control-cost-based frameworks by taking stochastic action as the primary trajectory-level organizing quantity, while treating those other quantities as possible contributors to the action functional, constraints on the ensemble, or parameter dynamics such as changes in \(\beta(t)\), \(\Theta(t)\), or \(\Omega(t)\). In this sense, the framework connects the variational structure of action-based physics to measurable ensemble-level diagnostics of organization, including \(\langle I\rangle_t\), \(\mathrm{Var}_t[I]\), and derived efficiency measures.

The thermodynamic interpretation of the stochastic action in dissipative systems is that $I[\Gamma]$ serves as the fundamental cost functional of the ensemble. Within dissipative systems, $I[\Gamma]$ is often associated with cumulative dissipation and the temporal duration of the realization \citep{seifert2012stochastic,touchette2009large}. Consequently, the ensemble-level evolution of $\langle I\rangle_t$ described by \Cref{cor:precision} reflects systematic shifts in the distribution of dissipation-weighted trajectories rather than stochastic fluctuations of individual paths. The evolution identity yields testable constraints through ensemble observables such as $\langle I\rangle_t$ and $\mathrm{Var}_t[I]$, which may be estimated at either the microscopic or a suitably coarse-grained level. Since $\langle I\rangle_t$ has the dimensions of action, however, its numerical value is not directly comparable across different physical systems.

For this reason, the Average Action Efficiency (AAE) can be used as a derived dimensionless diagnostic. It is defined whenever the ensemble-average action is positive, \(\langle I\rangle_t>0\), as an inverse monotone of \(\langle I\rangle_t\). Thus, lower ensemble-average action corresponds to higher AAE, making \(\alpha_t\) a scale-independent indicator of action-based organization. This definition is not restricted to the precision-driven regime: changes in \(\alpha_t\) reflect the total change in \(\langle I\rangle_t\), whether that change arises from selectivity modulation, structural deformation, or support evolution. In the fixed-support setting, its evolution follows from the full AAP identity, while evolving admissible support would contribute additional support-dependent terms through the corresponding change in \(\langle I\rangle_t\). In the precision-driven regime, AAE obeys the closed sign relation \(\operatorname{sign}(\dot{\alpha}_t)=\operatorname{sign}(\dot{\beta}(t))\). When the observed evolution of \(\alpha_t\) cannot be accounted for by this precision-driven relation, beyond statistical or model-identification uncertainty, the discrepancy may instead indicate structural deformation, support evolution, poor identification of the effective variables, or breakdown of the canonical precision-driven approximation. These relations provide falsifiable constraints on the precision-driven regime. 

Beyond aggregate scaling and sign tests, the framework also imposes consistency requirements on the action functional itself: for example, the evolution law is invariant under additive shifts \(I[\Gamma]\rightarrow I[\Gamma]+C\), since such shifts leave \(\mathrm{Var}_t[I]\) unchanged. Therefore, any model in which a mere baseline shift of the action alters \(\dot{\langle I\rangle}_t\), with trajectory differences unchanged, alters the predicted precision-driven evolution fails to identify the action functional relevant to the canonical path-ensemble description.

A distinct diagnostic concerns the onset of structural deformation. If independent evidence indicates substantial changes in the action structure \(\Theta(t)\), but \(\dot{\langle I\rangle}_t\) remains accurately described by the precision-driven law, then the proposed structural variables are not contributing measurably to the ensemble-level dynamics. They may be dynamically irrelevant, already absorbed into the effective action, or incorrectly identified.

The precision-driven evolution law also admits an information-geometric interpretation. For canonical path ensembles of exponential form, the associated statistical model defines an exponential family whose Fisher--Rao metric is given by the covariance of the sufficient statistics \citep{amari2000methods}. In the one-parameter precision-driven regime, this covariance reduces to the variance of the stochastic action, so $\mathrm{Var}_t[I]$ acts both as the dynamical susceptibility in \Cref{cor:precision} and as the local metric factor controlling motion on the statistical manifold. In this setting, the variance-controlled evolution of $\langle I \rangle_t$ can be viewed as inducing a trajectory on the statistical manifold, with a rate controlled by the local statistical metric. When structural parameters $\Theta$ also evolve, the geometry becomes multidimensional and the dynamics involve the full covariance tensor rather than a single variance term.

\begin{table}[t]
\centering
\caption{Extensions of the action-weighted path-ensemble formulation relative to established static or asymptotic formulations.}
\label{tab:distinction}
\renewcommand{\arraystretch}{1.2}
\begin{tabular}{@{}p{3cm} p{4.5cm} p{5cm}@{}}
\toprule
\textbf{Aspect} & \textbf{Classical formulations} & \textbf{Present formulation} \\
\midrule
Selectivity \((\beta)\) & Fixed inverse-noise/constraint & Time-dependent selectivity \(\beta(t)\) \\
Action structure & Fixed action functional & Time-dependent action structure \(I_{\Theta(t)}[\Gamma]\) \\
Admissible support & Fixed path class & Fixed in the main theorem; evolving \(\Omega(t)\) gives additional boundary/transport terms \\
Time interpretation & Static, asymptotic, or parametrically specified & Physical-time evolution of the path ensemble \\
Ensemble-level law & No general physical-time evolution law for \(\langle I\rangle_t\) & Exact fixed-support evolution identity; closed variance-controlled law in the precision-driven regime\\
\bottomrule
\end{tabular}
\end{table}

The current approach connects to frameworks that provide action functionals, inference principles, thermodynamic accounting, control methods, guidance mechanisms, adaptive-bias mechanisms, or steady-state selection principles. Information thermodynamics, dissipation-driven adaptation, Guided Self-Organization, the Free Energy Principle, and control theory can generate effective trajectory bias, precision, or selectivity through mechanisms, constraints, protocols, or interpretive models \citep{sagawa2010generalized,horowitz2014second,england2013statistical,Prokopenko2009GSO,Friston2010UnifiedFEP,kappen2005path}. The role of the AAP is to describe the ensemble-level consequence once such mechanisms alter trajectory weights, action structure, or admissible histories. In this sense, the AAP translates mechanisms supplied by existing theories into measurable changes in action-weighted trajectory organization within a variational path-space formulation. 

The most direct mathematical interfaces occur with established path-space formulations. When \(I[\Gamma]\) coincides with the Onsager--Machlup functional, the canonical ensemble recovers the corresponding stochastic path weighting in the linear Gaussian regime \citep{machlup1953fluctuations,seifert2012stochastic,graham1977covariant}. In the weak-noise limit at fixed precision, it recovers the Freidlin--Wentzell large-deviation structure \citep{freidlin1998random,touchette2009large}. At fixed constraints, the same exponential form is also the Maximum Caliber solution obtained by maximizing path entropy subject to an action constraint \citep{jaynes1980minimum,presse2013principles,dixit2018MaxCal}. These correspondences identify the static or asymptotic path-ensemble forms on which the present physical-time response law builds. In this way, inference-based path ensembles and feedback-driven dynamics can be understood within a common structural setting without modifying the underlying variational foundation. The key structural differences between classical stochastic variational formulations and the present framework are summarized in Table~\ref{tab:distinction}. 

Once a mechanism produces effective changes in selectivity, action structure, or admissible support, the AAP constrains how the action-weighted path ensemble reorganizes, within the limits of the fixed-support theorem and its possible extensions. Thus, the framework complements MEPP, GSO, dissipation-driven adaptation, the Free Energy Principle, and control theory by providing the path-space dynamical layer in which their mechanisms can be translated into measurable changes in \(\langle I\rangle_t\), \(\mathrm{Var}_t[I]\), and AAE. The comparison  in Table~\ref{tab:interfaces}  should therefore be understood as a separation of explanatory levels.

Traditional information thermodynamics typically accounts for feedback through information-theoretic correction terms to entropy production, work, or thermodynamic inequalities \citep{sagawa2010generalized,sagawa2012nonequilibrium,horowitz2014second}. The present formulation addresses a complementary question: how does feedback redistribute probability over admissible action-bearing trajectories? Related path-space frameworks often assign quantities such as entropy production, work, or information flow to trajectories \citep{seifert2012stochastic,endres2017entropy}; the present formulation instead takes stochastic action as the central trajectory-level organizing quantity, grounded in the variational role of action in mechanics, path-integral theory, Onsager--Machlup theory, and large-deviation descriptions. 

At each instant, the static canonical path measure in \Cref{eq:SDLAP_base} coincides with the MaxCal solution by maximizing path entropy subject to a constraint on the instantaneous expectation of the action \citep{jaynes1980minimum,presse2013principles}. Maximum Caliber therefore supplies the least-biased path ensemble at fixed constraints, but it does not by itself provide a physical-time law for how that ensemble changes when selectivity or action structure evolves. The present formulation adds this dynamical layer by treating the canonical path ensemble as a time-dependent object. For fixed \(\Theta\) and fixed admissible support, strengthening nonnegative selectivity, \(\dot{\beta}>0\), also decreases the path entropy of the canonical weights while the ensemble remains maximally entropic at each instant relative to its current constraint. Thus, Maximum Caliber identifies the instantaneous canonical ensemble, whereas the AAP describes how the system moves through the family of such ensembles during organization.

The current framework complements thermodynamic descriptions such as dissipation-driven adaptation \citep{england2013statistical, england2015dissipative}. While England argues that directionality arises from forward--backward path asymmetries (e.g., via the Crooks fluctuation theorem), the formalism here derives directional evolution from the kinematic deformation of the measure via $\beta(t)$. From this perspective, England’s thermodynamic driving is one possible physical mechanism capable of increasing effective path-space selectivity. This separates the kinematic response of the canonical ensemble from the specific thermodynamic asymmetries that may drive it. In the precision-driven case, this connects fluctuation-theorem-based accounts of adaptive bias with the variance-controlled AAP law for transient path-ensemble organization. More generally, thermodynamic asymmetries may also alter the action structure \(\Theta(t)\), but those effects enter through the structural terms of the full AAP identity rather than through the closed variance-controlled branch.

Guided Self-Organization (GSO) emphasizes how constraints, information flows, embodied feedback, and external or hybrid guidance can steer the emergence of organized behavior without prescribing every microscopic degree of freedom \citep{Prokopenko2009GSO,AyDerProkopenko2012GSO,ProkopenkoPolaniAy2014CrossDisciplinary,ProkopenkoGershenson2014EntropyMethodsGSO,PolaniProkopenkoYaeger2013InfoSelfOrgBehavior}. The present formulation addresses a complementary question: once such guidance changes trajectory preferences, how does the full path ensemble respond? In the action-based path-ensemble description, guidance can be represented as modulation of selectivity \(\beta(t)\), deformation of the action structure \(\Theta(t)\), or, more generally, changes in the admissible trajectory set \(\Omega(t)\). This representation is useful because it converts guidance into measurable changes in \(\langle I\rangle_t\), action variance, and AAE within an action-based variational path-space formulation. This allows one to distinguish whether guidance concentrates probability toward lower-action trajectories, shifts weight toward higher-action trajectories, or requires structural or support effects beyond the precision-driven approximation. Thus, GSO and the present framework address different levels of the same problem: GSO concerns the origin or design of guidance mechanisms, whereas the AAP gives an action-based response law for their effect on trajectory statistics.

The relation to stochastic and path-integral control is important because both frameworks use trajectory-level costs and, in some formulations, exponential weighting of paths \citep{leith2000survey,aastrom1995adaptive,kappen2005path}. In optimal control, however, the central problem is to design a protocol, policy, or feedback law that steers a system toward a prescribed objective by minimizing an expected cost. The present formulation addresses a different question: once selectivity, action structure, or admissible support changes, how does the probability distribution over action-bearing trajectories reorganize in physical time? External controllers may generate such changes, but they are not required; endogenous feedback can play the same formal role. This is useful for self-organizing systems because externally controlled and internally generated changes can be represented in the same path-ensemble variables, while remaining distinguished by their origin. In the precision-driven regime, the consequence of such modulation is constrained, giving a measurable response law rather than a prescription for an optimal policy.

\begin{table*}[!htbp] 
\centering
\footnotesize
\caption{Established frameworks and the distinct contribution of the present action-based path-ensemble formulation. The comparison is intended to identify levels of description:
existing frameworks may provide feedback mechanisms, inference principles, thermodynamic accounting, guidance or control protocols, that affect  $\beta(t)$, $\Theta(t)$ and $\Omega(t)$, while the present work provides an exact physical-time response law for action-weighted trajectory ensembles.}
\label{tab:interfaces}
\renewcommand{\arraystretch}{1.25}
\begin{tabular}{@{}p{1.6cm}p{2.5cm}p{3.5cm}p{5cm}@{}}
\toprule
\textbf{Framework} &
\textbf{Main organizing quantity or level} &
\textbf{Primary strength of the established approach} &
\textbf{Distinct contribution of the present work} \\
\midrule

Onsager--Machlup / large deviations &
Stochastic action, rate functions, most probable paths, rare fluctuations &
Provides action or rate-function descriptions of trajectory probabilities for specified stochastic dynamics, especially in small-noise, asymptotic, or fluctuation regimes. &
Shares the action-based path-space foundation, but promotes the canonical path ensemble itself to a physical-time evolving object. \\

Maximum Caliber &
Path entropy and constraints on trajectory observables &
Infers path distributions by maximizing path entropy subject to dynamical constraints; supplies an instantaneous or constrained inference principle for path ensembles. &
Uses the resulting canonical path-ensemble form as a dynamical object rather than only an inferred static distribution. The AAP describes physical-time evolution of the ensemble when \(\beta(t)\) and \(\Theta(t)\) vary. \\

Stochastic thermodynamics / fluctuation theorems &
Work, heat, entropy production, irreversibility along stochastic trajectories &
Provides thermodynamic accounting at the trajectory level and exact fluctuation relations for nonequilibrium processes.&
Thermodynamic quantities may enter through \(I_\Theta[\Gamma]\), through effective selectivity \(\beta(t)\) when noise, temperature, driving, or feedback precision changes, or through boundary/source--sink constraints that affect \(\Omega\).\\

Information / feedback thermodynamics &
Information flow, measurement, feedback, work extraction, entropy-production corrections &
Explains how measurement and feedback modify thermodynamic inequalities, work extraction, and entropy-production balances. &
Represents feedback as a possible modulation of path-space selectivity \(\beta(t)\) or action structure \(\Theta(t)\). This turns feedback into a calculable change in trajectory weights and predicts its effect on ensemble-average action. \\

MEPP &
Entropy production, fluxes, steady-state or stationary nonequilibrium selection &
Addresses selection of nonequilibrium steady states or flux configurations under thermodynamic constraints. &
Places MEPP-type stationary selection within a broader transient path-ensemble setting. The AAP describes how action-weighted ensembles approach, leave, or reorganize between regimes when selectivity or action structure changes in physical time.\\

Dissipation-driven adaptation &
Dissipated work, thermodynamic asymmetry, adaptive trajectory bias &
Explains how driven many-body systems can acquire adaptive trajectory bias through thermodynamic asymmetries, providing one possible mechanism for increased effective selectivity. &
Represents such bias, when appropriate, as modulation of \(\beta(t)\). Once this representation is made, the AAP gives the ensemble-level response; in the precision-driven regime, the response is variance-controlled.\\

Guided Self-Organization &
Guidance mechanisms, constraints, information flows, embodied or external feedback &
Develops conceptual and design principles for steering emergent organization without prescribing every microscopic degree of freedom. &
Converts guidance into path-ensemble variables: selectivity modulation \(\beta(t)\), action-structure deformation \(\Theta(t)\), and support change \(\Omega(t)\). Organizational change is then quantified through \(\langle I\rangle_t\), \(\mathrm{Var}_t[I]\), and AAE.\\

Free Energy Principle &
Variational free energy, generative models, inferential precision, adaptive action &
Models perception, learning, and action through precision-weighted inference in adaptive systems. &
Distinguishes inferential precision in generative models from effective path-space selectivity \(\beta(t)\). When FEP-style precision regulation can be represented as modulation of \(\beta(t)\), the AAP gives a complementary response law for changes in action-bearing trajectory statistics, including \(\langle I\rangle_t\), \(\mathrm{Var}_t[I]\), and AAE.\\

Control theory / path-integral control &
Control costs, policies, protocols, steering of states or trajectory distributions &
Provides methods for designing protocols that steer systems toward desired behavior under cost functionals. &
Treats the evolving path ensemble as a physical object of description, not only as an optimization target. Control protocols or closure laws for \(\beta(t)\) and \(\Theta(t)\) can be analyzed through their effects on average action, fluctuations, and efficiency.\\

\bottomrule
\end{tabular}
\end{table*}

In Free Energy Principle-based formulations, precision plays a central role in inferential dynamics, learning, and action \citep{Friston2010UnifiedFEP,Friston2023FEP}. These approaches primarily address how adaptive systems estimate and regulate precision within generative models of perception and behavior. The present formulation treats precision in a different but potentially compatible statistical-mechanical setting: as the parameter \(\beta(t)\) controlling trajectory weighting in a canonical path ensemble. Thus, inferential precision and physical path-space selectivity should not be identified automatically, although they may be coupled in adaptive systems. FEP may provide mechanistic models for how precision-like variables are generated and regulated, while the AAP describes the ensemble-level consequence once such variables are represented as changes in effective path-space selectivity. In the precision-driven regime, this consequence is fixed by \cref{cor:precision}, making changes in trajectory organization measurable through \(\langle I\rangle_t\), \(\mathrm{Var}_t[I]\), and AAE. Thus, when FEP-style precision regulation can be represented as modulation of effective path-space selectivity, the AAP supplies a complementary response law for how that modulation changes the distribution of action-bearing trajectories.

The relation to MEPP is best understood as a limiting-case relation. When \(\beta\), \(\Theta\), and the admissible support \(\Omega\) are fixed, the canonical path measure is stationary at the ensemble level; it corresponds to a physical nonequilibrium steady state only under the additional NESS correspondence discussed above. Subject to this correspondence, MEPP-type results may arise as limiting steady-state selection principles in stationary source--sink settings with fixed boundary fluxes \citep{dewar2003information,dyke2010maximum,georgiev2025Bio1}. The present framework addresses the complementary transient question: how does an action-weighted path ensemble approach, leave, or reorganize between such regimes when selectivity or action structure changes in physical time? Thus, MEPP concerns steady-state or stationary-flux selection, whereas the AAP gives a dynamical response law for changes in \(\langle I\rangle_t\), including the precision-driven identity in \Cref{cor:precision}.

Future work will focus on identifying conditions under which structural and support evolution, despite their generally system-dependent character, admit reduced or approximately closed descriptions, and on characterizing how such dynamics couple to precision modulation in multi-parameter feedback systems. In particular, situations in which $\beta(t)$, $\Theta(t)$ and $\Omega(t)$ co-evolve suggest the emergence of higher-order feedback mechanisms governing large-scale organization, adaptation, and innovation in complex systems. Additional directions include extending the formalism to non-smooth or discontinuous parameter evolution, where the differential identity must be replaced by integral formulations incorporating jump contributions. Understanding how external perturbations and transient disturbances interact with feedback-controlled selectivity may clarify how precision-driven regimes are initiated, stabilized, or disrupted in realistic systems, and may help delineate the conditions under which organized behavior emerges and persists under realistic, time-dependent driving.

\vspace{6pt}

\section{Methods}

\subsection{Admissible trajectories and event definition}
\label{MethodsOmega}

The variational framework is formulated in path space and is not restricted to source--sink first-passage events. In the open-system setting emphasized here, admissible trajectories are taken to be source--sink realizations representing systems sustained by fluxes. Let \(\Sigma_{\mathrm{src}}\) and \(\Sigma_{\mathrm{sink}}\) denote fixed source and sink manifolds in the system's configuration space. The chosen event definition restricts the full path space to an admissible set \(\Omega\), corresponding here to first-passage trajectories from \(\Sigma_{\mathrm{src}}\) to \(\Sigma_{\mathrm{sink}}\):
\begin{equation}
\Omega \equiv \left\{ \Gamma : \Gamma(0) \in \Sigma_{\mathrm{src}}, \quad \tau_{\mathrm{sink}}(\Gamma) < \infty, \quad \Gamma\left(\tau_{\mathrm{sink}}(\Gamma)\right) \in \Sigma_{\mathrm{sink}} \right\},
\end{equation}
where $\tau_{\mathrm{sink}}(\Gamma)$ denotes the first-hitting time of the sink manifold. This restriction ensures that the ensemble is composed of completed functional cycles, allowing the stochastic action $I[\Gamma]$ to be evaluated over well-defined realizations of system activity. The sink may be non-absorbing, allowing continued evolution, or absorbing with renewal or sustained driving that reinitializes trajectories at the source. Each trajectory represents a completed event and the path measure remains normalizable \citep{redner2001guide}.

\subsection{Stochastic--dissipative action functional}
\label{MethodsActionFunctional}

Each admissible trajectory \(\Gamma \in \Omega\) is assigned a stochastic--dissipative action functional \(I_{\Theta}[\Gamma]\). The parameters \(\Theta\) specify the structure of this functional, such as dynamical coefficients, fields, geometric constraints, or other quantities that determine how trajectories are evaluated within the ensemble. On intervals where \(\dot{\Theta}=0\), the action functional has no explicit parametric time dependence, and we write \(I[\Gamma]\) for simplicity.

The functional form of \(I_{\Theta}[\Gamma]\) is system-dependent. For Markov diffusion processes, it may be identified with an Onsager--Machlup action functional \citep{machlup1953fluctuations}; more general large-deviation formulations provide action or rate-function descriptions for broader stochastic dynamics \citep{freidlin1998random,touchette2009large,graham1977covariant,seifert2012stochastic}. In this paper, \(I_{\Theta}[\Gamma]\) is used as a trajectory-level cost functional whose parametric dependence on \(\Theta\) permits the structural-deformation terms in \Cref{Theorem1}.

\subsection{Static canonical ensemble and least-action selective regime}

The static canonical path ensemble used in \Cref{eq:SDLAP_base} is the fixed-parameter limit of \Cref{eq:Canonical_dynamic}, obtained by holding \(\beta\), \(\Theta\), and \(\Omega\) fixed:
\[
P[\Gamma]=\frac{1}{Z}\exp[-\beta I_\Theta[\Gamma]],
\qquad
Z=\int_{\Omega}\exp[-\beta I_\Theta[\Gamma]]\,\mathcal{D}\Gamma.
\]
At the formal level, \(\beta\) is any admissible parameter for which the canonical path ensemble remains normalizable. In the physical interpretation used here, the relevant regime is \(\beta\ge 0\). For \(\beta>0\), lower-action trajectories are selectively favored, and in the large-selectivity limit the ensemble concentrates on minimizers, or minimizing sets, of the stochastic action \citep{freidlin1998random,touchette2009large,gardiner2009stochastic}. This positive-selectivity static limit corresponds to the stochastic--dissipative least action principle (SDLAP) \citep{georgiev2025Bio1}.

The boundary case \(\beta=0\) gives uniform weighting over admissible trajectories; negative \(\beta\) formally reverses the least-action ordering and is not used in the physical interpretation adopted here.

\subsection{Assumptions and Domain of Validity}
\label{MethodsAssumptions}

Throughout the derivation, the ensemble is assumed to have the canonical form in \Cref{eq:Canonical_dynamic}, with stochastic action \(I_{\Theta(t)}[\Gamma]\), time-dependent parameters \(\beta(t)\) and \(\Theta(t)\), and fixed admissible support \(\Omega\) on each interval where the differential identity is applied. The core evolution identity is an ordinary differential relation valid piecewise on intervals where \(\Omega\) is fixed and where \(\beta(t)\) and \(\Theta(t)\) are sufficiently regular. Smooth structural deformation through \(\Theta(t)\) is included within these intervals, whereas support evolution \(\Omega(t)\) requires additional support-dependent boundary or transport terms and is treated separately.

The ensemble-level dynamical results require that the canonical path ensemble in \Cref{eq:Canonical_dynamic} be well defined and normalizable, with finite and nonzero partition functional \(Z_t\) on each interval considered. Signed or reference-shifted stochastic--dissipative action functionals are admissible under the same condition. On each interval considered, the canonical measure is assumed to possess finite first and second moments of the action, finite expectation of \(\partial_\Theta I_\Theta\), and finite covariance \(\mathrm{Cov}_t(I_\Theta,\partial_\Theta I_\Theta)\).

The deformation parameters \(\beta(t)\) and \(\Theta(t)\) are assumed to be continuously differentiable on intervals where ensemble evolution is considered, and the stochastic action is differentiable with respect to \(\Theta\) wherever parametric derivatives appear. The parametric derivatives \(\partial_\Theta I_\Theta[\Gamma]\) are assumed to be integrable with respect to the canonical measure. Under these conditions, ensemble averages are differentiable in time and admit the exact decomposition stated in \Cref{Theorem1}. Differentiation under the path integral follows from standard dominated-convergence arguments, provided that the stochastic action and its parametric derivatives satisfy the corresponding integrability and domination conditions \citep{gardiner2009stochastic,risken1989fokker}.

The kinematic identity holds for any sufficiently regular time-dependent \(\beta(t)\) for which the canonical path ensemble remains well defined and normalizable. Mathematically, \(\beta(t)\) enters only as the time-dependent multiplier conjugate to the stochastic action in the exponential path weight. The least-action selective interpretation enters only in the nonnegative regime, \(\beta(t)\ge 0\), where increasing \(\beta(t)\) strengthens the relative weighting of lower-action trajectories. This is the physical convention adopted in the paper. The results apply only to systems admitting a well-defined trajectory-level description and a canonical path-ensemble representation.

\subsection{Log-derivative identities}

\begin{lemma}[Log-derivative identity for fixed admissible support]
\label{lem:path_weight}
On intervals where the admissible trajectory set $\Omega(t)$ is fixed, the canonical path measure
\[
P_t[\Gamma]
=
\frac{1}{Z_t}\exp[-\beta(t)I_{\Theta(t)}[\Gamma]]
\]
satisfies
\[
\frac{d}{dt}\log P_t[\Gamma]
=
-\dot{\beta}(t)\left(I_{\Theta(t)}[\Gamma]-\langle I\rangle_t\right)
-\beta(t)\left(\partial_\Theta I_{\Theta(t)}[\Gamma]-\langle \partial_\Theta I_\Theta\rangle_t\right)\cdot \dot{\Theta}(t),
\]
provided the regularity and integrability assumptions stated in \nameref{MethodsAssumptions} hold.
\end{lemma}

\begin{proof}[Proof sketch]
The identities follow by direct differentiation of the canonical path measure \Cref{eq:Canonical_dynamic}. Differentiating $\ln Z_t$ yields
\[
\partial_t \ln Z_t
=
\frac{\partial_t Z_t}{Z_t}
=
\frac{1}{Z_t}\int_{\Omega} \left[-\dot{\beta}(t)I_\Theta[\Gamma]-\beta(t)\partial_\Theta I_\Theta[\Gamma]\cdot\dot{\Theta}(t)\right] e^{-\beta I_\Theta[\Gamma]}\,\mathcal{D}\Gamma,
\]
which gives the stated expression in terms of ensemble averages. Similarly, using $\partial_t \ln P_t = \partial_t(-\beta I_\Theta - \ln Z_t)$ and substituting the expression for $\partial_t \ln Z_t$ yields the second identity.
\end{proof}

\subsection{Proof of Theorem 1}
\label{Methods:Proof}
\begin{proof}[Theorem Proof sketch]

Differentiating the ensemble average yields
\begin{equation}
\dot{\langle I\rangle}_t
=
\int_{\Omega} I_{\Theta(t)}[\Gamma]\,\partial_t P_t[\Gamma]\,\mathcal{D}\Gamma
+
\int_{\Omega} (\partial_t I_{\Theta(t)}[\Gamma])\,P_t[\Gamma]\,\mathcal{D}\Gamma.
\end{equation}

Using $\partial_t P_t = P_t\,\partial_t \ln P_t$ and substituting the expression for $\partial_t \ln P_t$ from Lemma~\ref{lem:path_weight}, the first term becomes
\begin{align}
\int_{\Omega} I_\Theta[\Gamma]\,\partial_t P_t[\Gamma]\,\mathcal{D}\Gamma
&=
\big\langle I_\Theta\,\partial_t \ln P_t \big\rangle_t \\
&=
-\dot{\beta}(t)\,\mathrm{Var}_t[I]
-\beta(t)\,\mathrm{Cov}_t\!\big(I_\Theta,\partial_\Theta I_\Theta\big)\cdot\dot{\Theta}(t).
\end{align}

The second term yields
\begin{equation}
\int_{\Omega} (\partial_t I_{\Theta})\,P_t\,\mathcal{D}\Gamma
=
\left\langle \partial_\Theta I_\Theta \right\rangle_t \cdot \dot{\Theta}(t).
\end{equation}

Combining both contributions yields the stated identity.

\end{proof}

\vspace{6pt}

\paragraph{Data availability: }No data were used in this work. 

\paragraph{Competing interest.} The author is guest editor of npj Complexity, "Variational, Nonequilibrium, and Optimization Principles of the Coevolution of Structure and Dynamics in Complex Systems". The author was not involved in the journal’s review of, or decisions related to, this manuscript.

\paragraph{Author contributions.} This is a single-author article; all work was performed by the author.

\paragraph{Acknowledgments.} The author acknowledges support from Assumption University through Faculty Development and Course Load Reduction grants, as well as support for undergraduate summer research. Additional institutional support was provided by Worcester Polytechnic Institute. No external funding.

\bibliography{main, references}

\end{document}